\documentclass[fleqn,10pt]{wlscirep}
\usepackage[utf8]{inputenc}
\usepackage[T1]{fontenc}
\usepackage{bm}
\title{Polarization-sensitive PHASR Scanner and calibration technique for accurate mapping of the Stokes vectors in terahertz frequencies}

\author[1]{Zachery B. Harris}
\author[1]{Kuangyi Xu}
\author[1,*]{M. Hassan Arbab}
\affil[1]{Stony Brook University, Biomedical Engineering Department, Stony Brook, NY, USA}

\affil[*]{hassan.arbab@stonybrook.edu}


\begin{abstract}
In recent years, handheld and portable terahertz instruments have been in rapid development for various applications ranging from non-destructive testing to biomedical imaging and sensing. For instance, we have deployed our Portable Handheld Spectral Reflection (PHASR) Scanners for \emph{in vivo} full-spectroscopic imaging of skin burns in large animal models in operating room settings. In this paper, we debut the polarimetric version of the PHASR Scanner, and describe a generalized calibration technique to map the spatial and spectral dependence of the Jones matrix of an imaging scanner across its field of view. Our design is based on placement of two orthogonal photoconductive antenna (PCA) detectors separated by a polarizing beam splitter in the PHASR Scanner housing. We show that at least three independent measurements of a well-characterized polarimetric calibration target are sufficient to determine the polarization state of the incident beam at the sample location, as well as to extract the Jones propagation matrix from the sample location to the detectors. We have tested the accuracy of our scanner by validating polarimetric measurements obtained from a birefringent crystal rotated to various angles, as compared to the theoretically predicted response of the sample. This new version of our PHASR scanner can be used for high-speed imaging and investigation of heterogeneity of polarization-sensitive samples in the field.

\end{abstract}
\begin{document}

\flushbottom
\maketitle

\thispagestyle{empty}

\section*{Introduction}

Novel polarimetric and ellipsometry applications in the terahertz (THz) frequency range have received significant attention in recent years\cite{leitenstorfer20232023, chen2022introduction, nagashima2013polarization}. Several industrial applications have been demonstrated including characterization of materials such as polymer composites \cite{okano2016anisotropic, okano2017anisotropic}, semiconductors\cite{mittleman1997noncontact, nagashima2001measurement, shuvaev2011giant}, and metamaterials\cite{singh2010highly, zhou2012terahertz}. Polarimetric measurements have also provided an accurate way to characterize the properties of thin films\cite{matsumoto2011measurement, neshat2012terahertz}, detect fiber orientation\cite{katletz2012polarization}, and map internal strain of polymers \cite{moriwaki2017internal}.
Additionally, biological samples present rich and complex opportunities in investigating the structure and function of proteins in biomolecule crystals\cite{acbas2014optical, niessen2019protein, deng2021near}. The change in polarization of THz light due to scattering from inhomogeneities in biological tissue provides insights into tissue organization and can help identify cancerous areas in \emph{ex vivo} samples \cite{doradla2013detection, joseph2014imaging}. Similarly, the internal structures of skin produce characteristic spectroscopic and polarimetric signal contrast mechanisms \cite{Arbab:11:conf, Arbab:13, xu2024terahertz}. For instance, the polarimetric effect of the stratum corneum's stratified structure and its surface grooves on measurement of skin hydration have been quantified using an \emph{in vivo} THz ellipsometer \cite{chen2021exploiting}.



Some of these polarization-measurement techniques take advantage of the inherent sensitivity of the EO-sampling technique to the polarization direction of the THz waves \cite{van2005terahertz, yasumatsu2014high}. Alternatively, multi-contact photoconductive detectors have been developed, which simultaneously measure the orthogonal components of the electric field\cite{castro2005polarization, bulgarevich2014polarization}. Similarly, a network of spatially oriented nanowires have been used to achieve polarimetric measurements in THz frequencies \cite{peng2020three}. Other approaches rely on using external components such as wire grid polarizers\cite{chen:18, ikebe2008characterization, shimano2011terahertz, shimano2013quantum,chen2024real,mazaheri2022accurate}, optical prisms\cite{hofmann2010variable, shan2009circularly}, or waveplates\cite{jenkins2010simultaneous}. High-sensitivity methods have been developed by continuous rotation of wire-grid polarizers\cite{aschaffenburg2012efficient, george2012terahertz, morris2012polarization} or an EO-crystal\cite{xu2020terahertz, xu2022broadband} to achieve unprecedented accuracy or measurement speed.




With advancement of THz technology, high-speed and portable scanning systems have been introduced by multiple groups \cite{chen:22review, wilmink2011development, grootendorst2017use, hernandez2024terahertz, Harris:2020:IEEEAcc:PHASR1, virk2021, Virk2023}.
We have recently debuted our THz Portable HAndheld Spectral Reflection (PHASR) Scanner as a versatile instrument for high-speed broadband time-domain spectroscopic imaging of 1-inch scenery at up to 2.6 kHz trace acquisition rate in about 6-8 seconds \cite{Harris:2022:IEEETrans:2kHz}. We have demonstrated its use in a wide range of applications including \emph{in vivo} biomedical imaging \cite{Osman:22:boe, KhaniJBO22, Khani:2023:BOE:Triage}, non-destructive evaluation \cite{Harris:2020:IEEEAcc:PHASR1}, and spectral mapping through diffusive materials \cite{Khani:2022:OptExp:Multiresolution}. Despite these advancements, formation of full polarimetric images at high speeds such that it can be suitable for deployment in the field has been an illusive target. In this paper, we present a polarimetric version of our PHASR Scanner that can address these limitations by forming broadband THz images of the full Stokes vector of the light reflected from the target. In order to accurately describe the polarization state before and after the sample, we must account for optical (due to e.g., polarizers and beam splitter) and electronic (e.g., different PCA gain and coupling efficiency) effects that are introduced by the dual-channel system. Therefore, we describe a calibration technique that can determine both the polarization state of the incident light as well as the Jones matrix of the propagation path using only three independent measurements of a well-characterized calibration target, such as a wire grid polarizer, placed at the sample location. Using this technique, we demonstrate that the elements of the Jones matrix of the system and the incident polarization of the THz light both vary significantly over the field of the view of the scanner. Therefore, careful calibration of the imaging instrument, including the beam steering optics, are crucial in accurate imaging of polarization-sensitive targets and in studying heterogeneity of samples. Finally, we validate our approach by mapping the stokes vector of the THz light upon reflection from a birefringent crystal, i.e., sapphire, while the sample is rotated to various angles. Our results show excellent agreement to theoretically predicted response of the rotating crystal. 




\section*{Methods}
\subsection*{Jones matrix description of the instrument}
\begin{figure}[h]
\centering
\includegraphics[width=6.75in]{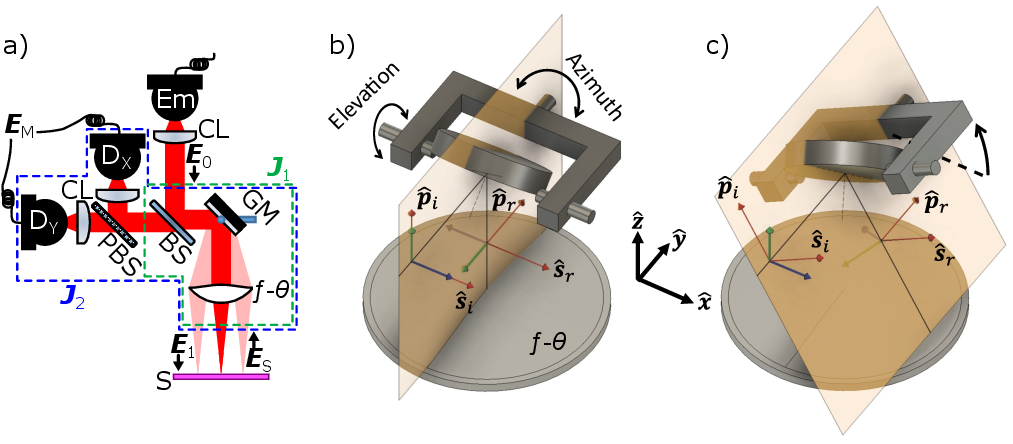}
\caption{\label{fig:1Schematic}\label{fig:1bcGimbalingandmirror} (\textbf{a}) A diagram of the polarization-sensitive THz PHASR Scanner 2.1. The THz light is generated by a PCA emitter (Em) while two PCA detectors ($\text{D}_{\text{X}}$ and $\text{D}_{\text{Y}}$) simultaneously measure the orthogonal polarization components of the THz field. Each of the PCAs are paired with a TPX lens for collimating/focusing the beam (CL). A silicon beam splitter (BS) directs the emitted THz light towards the motorized gimballed mirror (GM) for telecentric beam steering across an $f\text{-}\theta$ scanning lens, which focuses the light on the sample (S). The reflected THz light from the sample surface retraces the same optical path as the incident beam passing through the BS, after which its orthogonal polarization components are directed by a polarizing beam splitter (PBS) towards the PCA detectors. Dashed boxes indicate which components are included in $\bm{J}_{1}$ (green) and $\bm{J}_{2}$ (blue) jones matrices. 
The gimballed mirror causes a rotation of the polarization of the reflected light which is illustrated by comparing two cases: without mirror deflection (\textbf{b}) and deflected in the azimuthal direction (\textbf{c}). The orientation of the plane of incidence, and therefore $s$ and $p$ components (red arrows), change as the mirror steers the beam. A pair of orthogonal incident polarizations (green and blue) incident on the mirror show the effect at the location of the scanning lens (gray disk).}
\end{figure}

Figure~\ref{fig:1Schematic} illustrates the schematic and internal components of the polarization-sensitive version of the PHASR Scanner (i.e., version 2.1). The design of the previous versions of this scanner has been discussed elsewhere \cite{Harris:2020:IEEEAcc:PHASR1,Harris:2022:IEEETrans:2kHz}. In our nomenclature, the major version number of each generation of the PHASR Scanners (e.g., version 1.0, 2.0, etc.) are advanced according to the beam-steering methodology, the motion stage and controllers used internally. For example, PHASR 1.0 utilizes two linear stepper motors to steer the beam, whereas in version 2.0 a stacked combination of a rotation motor and a goniometer is used in a heliostat design. The decimal point numbers are reserved for optical/software advancements, such as acquisition speed, and number of detection channels. The PHASR Scanners are able to operate using either of ASOPS\cite{bartels2007ultrafast} or ECOPS\cite{tauser_electronically_2008} time-domain sampling techniques. To enable polarimetric capability, we use two photoconductive antenna (PCA) detectors oriented to measure orthogonal polarizations. The two detection PCAs are pumped by the same femtosecond laser, so any drift in the time domain is equally present in both channels. Furthermore, a wire-grid polarizer is used as a polarizing beam splitter (PBS) to direct orthogonal components of the THz field to each of the PCAs. The path of the THz light as it propagates through the system can be divided into three sections, each of which is described by a Jones matrix: the path from the emitter to the sample, $\bm{J}_{1}$; the response of the sample itself, $\bm{J}_{\text{S}}$; and finally the path from the sample to the detectors where the orthogonal components of the electric field are measured, $\bm{J}_{2}$. In this final matrix, we also include the intrinsic response function of the two PCAs, inclusive of any differences in detector sensitivity, alignment and electronics, as well as potential cross-talk between the channels. Thus, if the emitted THz light is expressed as the Jones vector $\bm{E}_{0}$, then the measured electric field vector $\bm{E}_{\text{M}}$, composed of the two signals detected by the orthogonal PCAs ($\text{D}_{\text{X}}$ and $\text{D}_{\text{Y}}$), is given by,
\begin{equation}
    \bm{E}_{\text{M}} = \bm{J}_{2}\bm{J}_{S}\bm{J}_{1}\bm{E}_{0}. 
    \label{eq:1BasicJones}
\end{equation}
Notably, the beam-steering components, i.e., the gimballed mirror (GM) and the $f\text{-}\theta$ scanning lens, are part of both the incident and return light paths, and must be accounted for when describing the polarization of the light at the sample and detectors.

\subsection*{Spatial variation of incident polarization due to the telecentric beam steering}
Figure~\ref{fig:1bcGimbalingandmirror}b-c illustrate the incidence polarization change due to telecentric beam steering in our system. The $s$ and $p$ polarization directions, indicated by the red arrows, are defined by the plane of incidence. Figure~\ref{fig:1bcGimbalingandmirror}b shows this plane and the polarization directions for an undeflected mirror targeting the center of the field-of-view on the sample. As the mirror is rotated about its azimuthal axis, the plane of incidence rotates with it, and so do the directions of the $s$ and $p$ components of the incident light, $\hat{\bm{s}}_{i}$ and $\hat{\bm{p}}_{i}$, as shown in Fig.~\ref{fig:1bcGimbalingandmirror}c. Upon reflection from the mirror, the direction of $s$ and $p$ polarized components of the reflected beam, $\hat{\bm{s}}_{r}$ and $\hat{\bm{p}}_{r}$, incident on the scanning lens (gray disk) will both depend on the azimuthal angle. In contrast, the rotation of the mirror along the elevation axis (not shown in Fig.~\ref{fig:1bcGimbalingandmirror}) will change the direction of $\hat{\bm{p}}_{r}$, but not $\hat{\bm{s}}_{r}$. Therefore, the polarization at the $f\text{-}\theta$ lens, and thus also at the sample location, changes over the field-of-view as the beam is steered \cite{Jepsen:2007Investigation}.

On the return path from the sample, the same effect rotates the beam polarization in the opposite direction. For isotropic targets, these rotations cancel out and are not observable in the measured data. However, in case of anisotropic samples, the effect of the variation of incident polarization over the field-of-view of the scanner will be significant. Therefore, both $\bm{J}_{1}$, and $\bm{J}_{2}$ are dependent on the angle of the gimbal and can impact polarimetric measurements. To capture this effect, and any potential spatial differences due to an imperfect alignment, a calibration procedure is performed for each pixel, as described below.

\subsection*{Measurement of the Jones matrices of the system}

To understand and calculate the polarimetric response of imaging targets, we must characterize the scanner in order to know: (I) the polarization state of the THz electric field immediately before the sample, and (II) how the polarization changes in the system between the sample and detectors. The field before the sample is calculated by $\bm{E}_{1}=\bm{J}_{1}\bm{E}_{0}$, while the changes after the sample are described by $\bm{J}_{2}$. With the substitution of $\bm{E}_{1}$ into Eq.~(\ref{eq:1BasicJones}), we express the system as, 
\begin{equation}
    \bm{E}_{\text{M}} = 
    \begin{bmatrix}
        E_{\text{M},x}\\
        E_{\text{M},y}\\
    \end{bmatrix} = \bm{J}_{2}\bm{J}_{S}\bm{E}_{1} = 
    \begin{bmatrix}
        s_{xx} & s_{xy}\\
        s_{yx} & s_{yy}\\
    \end{bmatrix}\bm{J}_{\text{S}}\begin{bmatrix}
        E_{1,x}\\
        E_{1,y}\\
    \end{bmatrix},
    \label{eq:2_E1ReplJones}
\end{equation}
where the expanded terms show the individual elements of $\bm{E}_{M}$, $\bm{J}_{2}$, and $\bm{E}_{1}$. Each of these elements are complex values that vary over the field-of-view (FOV) and frequency. Following this formulation, separate measurements of both $\bm{E}_{0}$ and $\bm{J}_{1}$ are not necessary as measurement of the the incident polarization on the sample, $\bm{E}_{1}$, is sufficient to fully characterize the system. However, direct mapping of $\bm{E}_{1}$ is impractical. Here, we describe a computational method that can extract the necessary calibration parameters by obtaining specific \emph{in situ} imaging scans of known polarimetric targets.

\begin{figure}[h]
\centering
\includegraphics[width=6.5in]{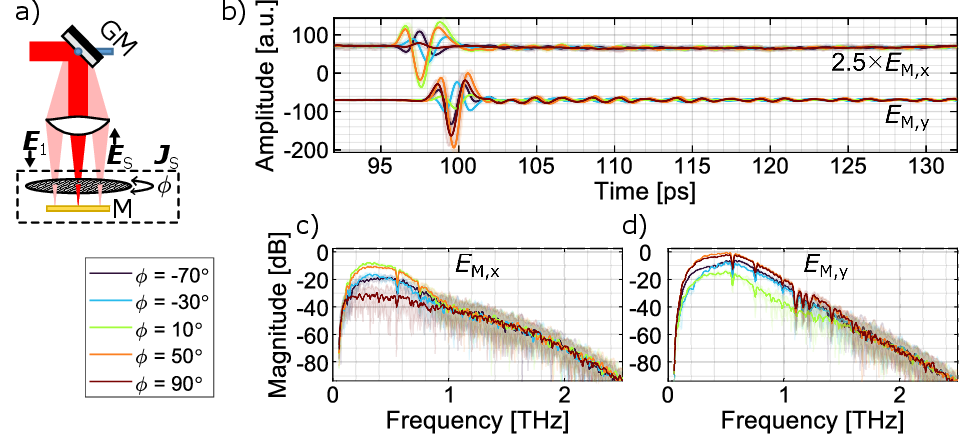}
\caption{\label{fig:2MirrorTarg} (\textbf{a}) Illustration of the mirror and polarizer sample used for calibration. The polarizer can be rotated to specified angles, $\phi$, to provide a set of well-described reflections from the mirror (M). When $\phi=0^\circ$, the polarizer is aligned to fully transmit $x$ polarized light. The black dashed box indicates the elements that constitute the sample Jones matrix, $\bm{J}_{\text{S}}$. (\textbf{b}) Terahertz time-domain trace  measurements (vertically offset), and (\textbf{c} and \textbf{d}) the corresponding amplitude spectra are shown for the $x$ and $y$ channels, respectively, at different polarizer angles. Shaded areas in each plot represent the full range of values measured along a $10\times1$ mm vertical line (ROI) through the center of the scanning area.}
\end{figure}

To simultaneously measure elements of both $\bm{J}_{2}$ and $\bm{E}_{1}$, we used a mirror at the target location, placed behind a wire-grid polarizer as illustrated in Fig.~\ref{fig:2MirrorTarg}a. The rotation angle of the polarizer when fully transmitting $x$ polarized light is defined to be $\phi=0^\circ$. The Jones matrix for the reflection from the mirror, passing twice through the polarizer rotated to angle $\phi$, is given by,
\begin{equation}
    \bm{J}_{\text{S}}(\phi) = \begin{bmatrix}
        -\cos^{2}(\phi) & -\cos(\phi)\sin(\phi)\\
        -\cos(\phi)\sin(\phi) & -\sin^{2}(\phi)\\
    \end{bmatrix}.
    \label{eq:3RotPolMat}
\end{equation} 
Different angles of the polarizer provide distinct sample response matrices. The measured polarization state is given by,
\begin{equation}
    \bm{E}_{M}(\phi) = \bm{J}_{2}\bm{J}_{S}(\phi)\bm{E}_{1}.
    \label{eq:4UnmodPolEq}
\end{equation} 

It should be noted that there are six unknown elements in Eq.~(\ref{eq:2_E1ReplJones}), and  each calibration measurement represents a series of two equations as a function of frequency (one for each of the $x$ and $y$ components). Thus, a sufficient calibration set consists of at least three measurements of different samples with known $\bm{J}_{\text{S}}$ matrices to provide six independent equations, which can fully define the unknown elements. Here, to obtain such a calibration set, we rotate the polarizer to a few different pre-determined angles. Figure~\ref{fig:2MirrorTarg}b-d illustrate the measured reflection from such a sample in time- and frequency-domains for a $10\times1$ mm vertical line of pixels at the center of the scan, where the polarization was not rotated by the gimballed mirror. 
The THz emitter was positioned at $45^\circ$ to nominally split $\bm{E}_{0}$ between the two detection orientations. Despite this, the field recorded by the $y$-component is approximately 2.5 times that of the $x$-component due to the internal components of the system itself, emphasizing the need for proper calibration. 


However, even for this well-characterized sample, Eq.~(\ref{eq:1BasicJones}) is not readily solvable for the six unknown values. This can be seen by expanding Eq.~(\ref{eq:4UnmodPolEq}) (not shown here), since the elements of $\bm{E}_{1}$ and $\bm{J}_{2}$ are not separable in a standard matrix factorization form. Here, we present an equivalent version that allows for matrix decomposition into a system of linear equations such that a vector composed of all of the independent (unknown) variables are multiplied by a matrix of known coefficients. We start by factoring the value of $s_{xx}$ in Eq.~(\ref{eq:2_E1ReplJones}) out of $\bm{J}_{2}$ and into a new vector, $\bm{E}'_{1}$, using the relationships,
\begin{equation*}
    \bm{E}_{1}' = \begin{bmatrix}
            E'_{1,x}\\
            E'_{1,y}\\
        \end{bmatrix} = s_{xx}\bm{E}_{1} =  s_{xx}\begin{bmatrix}
            E_{1,x}\\
            E_{1,y}\\
        \end{bmatrix} \quad \textrm{and} \quad \bm{J}'_{2} = \frac{\bm{J}_{2}}{s_{xx}} = \begin{bmatrix}
            1 & s_{1}\\
            s_{2} & s_{3}\\
        \end{bmatrix}
\end{equation*}
such that
\begin{equation}
    \bm{E}_{\text{M}} = \begin{bmatrix}
            1 & s_{1}\\
            s_{2} & s_{3}\\
        \end{bmatrix}\bm{J}_{\text{S}}\begin{bmatrix}
            E'_{1,x}\\
            E'_{1,y}\\
        \end{bmatrix}.
    \label{eq:5JonesPrime}
\end{equation}
This form reduces the polarimetric effects of the system to the five complex values $E'_{1,x}$, $E'_{1,y}$, $s_{1}$, $s_{2}$, and $s_{3}$. Despite this reduction in the number of variables, the new form has no impact on either of $\bm{J}_{S}$ or $\bm{E}_{M}$. Therefore, the calibration using the five elements in Eq.~(\ref{eq:5JonesPrime}) is equivalent to using the six elements in Eq.~(\ref{eq:2_E1ReplJones}). The five unknown elements from $\bm{E}'_{1}$ and $\bm{J}'_{2}$ in Eq.~(\ref{eq:5JonesPrime}) are still not independent from each other. However, by applying Eq.~(\ref{eq:5JonesPrime}) to the mirror-polarizer calibration sample described by Eq.~(\ref{eq:3RotPolMat}) and fully expanding, we can express the measured polarization, $\bm{E}_{\text{M}}(\phi)$, using a more tractable refactored form,


\begin{equation}
    \begin{split}
        \bm{E}_{M}(\phi) &= \bm{J}'_{2}\bm{J}_{S}(\phi)\bm{E}'_{1}\\
        \begin{bmatrix}
            E_{\text{M},x}(\phi)\\
            E_{\text{M},y}(\phi)\\
        \end{bmatrix} &= -\begin{bmatrix}
            1 & s_{1}\\
            s_{2} & s_{3}\\
        \end{bmatrix}\begin{bmatrix}
            \cos^{2}(\phi) & \cos(\phi)\sin(\phi)\\
            \cos(\phi)\sin(\phi) & \sin^{2}(\phi)\\
        \end{bmatrix}\begin{bmatrix}
            E'_{1,x}\\
            E'_{1,y}\\
        \end{bmatrix}\\
        &= -\begin{bmatrix}
            1 & s_{1}\\
            s_{2} & s_{3}\\
        \end{bmatrix}\begin{bmatrix}
            E'_{1,x}\cos^{2}(\phi) + E'_{1,y}\cos(\phi)\sin(\phi)\\
            E'_{1,x}\cos(\phi)\sin(\phi) + E'_{1,y}\sin^{2}(\phi)\\
        \end{bmatrix}\\
        &= -\begin{bmatrix}
            E'_{1,x}\cos^{2}(\phi) + E'_{1,y}\cos(\phi)\sin(\phi) + 
            s_{1}E'_{1,x}\cos(\phi)\sin(\phi) + s_{1}E'_{1,y}\sin^{2}(\phi)\\
            s_{2}E'_{1,x}\cos^{2}(\phi) + s_{2}E'_{1,y}\cos(\phi)\sin(\phi) + 
            s_{3}E'_{1,x}\cos(\phi)\sin(\phi) + s_{3}E'_{1,y}\sin^{2}(\phi)\\
        \end{bmatrix}\\
        &= -\begin{bmatrix}
            E'_{1,x}\cos^{2}(\phi) + (E'_{1,y}
            + s_{1}E'_{1,x})\cos(\phi)\sin(\phi) 
            + s_{1}E'_{1,y}\sin^{2}(\phi)\\
            s_{2}E'_{1,x}\cos^{2}(\phi)
            + (s_{2}E'_{1,y} + s_{3}E'_{1,x})\cos(\phi)\sin(\phi) 
            + s_{3}E'_{1,y}\sin^{2}(\phi)\\
        \end{bmatrix}\\
        \begin{bmatrix}
            E_{\text{M},x}(\phi)\\
            E_{\text{M},y}(\phi)\\
        \end{bmatrix} &= -\begin{bmatrix}
            \cos^{2}(\phi) & \cos(\phi)\sin(\phi) & \sin^{2}(\phi) & 0 & 0 & 0\\
            0 & 0 & 0 & \cos^{2}(\phi) & \cos(\phi)\sin(\phi) & \sin^{2}(\phi)\\
        \end{bmatrix}\begin{bmatrix}
            E'_{1,x}\\
            E'_{1,y} + s_{1}E'_{1,x}\\
            s_{1}E'_{1,y}\\
            s_{2}E'_{1,x}\\
            s_{2}E'_{1,y} + s_{3}E'_{1,x}\\
            s_{3}E'_{1,y}\\
        \end{bmatrix}.        
    \end{split}
    \label{eq:6DeriveAuxVect}
\end{equation}
In Eq.~(\ref{eq:6DeriveAuxVect}), $\bm{E}_{\text{M}}$ is given by a linear system, where a $2\times6$ matrix composed of the known properties of the polarizer is multiplied by a $6\times1$ vector composed of the unknown calibration elements. We will call this six-element vector the auxiliary vector, and identify the individual elements as $A$ through $F$ given by, 
\begin{equation}
    \begin{bmatrix}
        A\\
        B\\
        C\\
        D\\
        E\\
        F\\
    \end{bmatrix} = \begin{bmatrix}
        E'_{1,x}\\
        E'_{1,y} + s_{1}E'_{1,x}\\
        s_{1}E'_{1,y}\\
        s_{2}E'_{1,x}\\
        s_{2}E'_{1,y} + s_{3}E'_{1,x}\\
        s_{3}E'_{1,y}\\
    \end{bmatrix}.
    \label{eq:7AuxVectLabels}
\end{equation}
The values of the auxiliary vector can be found using standard linear algebra techniques after measuring the signal at three or more polarizer angles. 
The five unknown calibration elements can be then determined based on the relationships between elements of the auxiliary vector as shown in Table~\ref{tab:Aux2Params}. Notably, while $E'_{1,x} = A$ and $s_{2} = D/A$ are uniquely determined by simple relationships, 
$s_{1}$ is given by the roots of a quadratic equation and is then used for the calculation of the values of $s_{3}$ and $E'_{1,y}$. The result is two sets of possible values for $s_{1}$, $s_{3}$, and $E'_{1,y}$. Therefore, in order to determine which root is the correct solution to the system of equations, the previously unused fifth element of the auxiliary vector, $E$, is compared to $s_{2}E'_{1,y} + s_{3}E'_{1,x}$ for both sets of solutions.

\begin{table}[ht]
\centering
\begin{tabular}{|c|c|c|}
\hline
Variable & Derivation & Number of solutions\\
\hline
$s_{1}$ & $\displaystyle\begin{aligned}\rule{0pt}{4ex}
        B &= \frac{C}{s_{1}} + s_{1}A \rightarrow{}
        s_{1} = \frac{B\pm\sqrt{(-B)^{2}-4AC}}{2A}\\[1ex]
\end{aligned}$ & 2 \\
\hline
$s_{2}$ & $\displaystyle \rule{0pt}{4ex}s_{2} = \frac{D}{E'_{1,x}} = \frac{D}{A}$ & 1\\[3ex]
\hline
$s_{3}$ & $\displaystyle \rule{0pt}{4ex}s_{3} = \frac{F}{E'_{1,y}} = \frac{F}{B - s_{1}A}$ & 2 \\[3ex]
\hline
\rule{0pt}{3ex}$E'_{1,x}$ & $\displaystyle E'_{1,x} = A$ & 1 \\[1ex]
\hline
\rule{0pt}{3ex}$E'_{1,y}$ & $\displaystyle E'_{1,y} = B - s_{1}E'_{1,x} = B - s_{1}A$& 2 \\[1ex]
\hline
\end{tabular}
\caption{\label{tab:Aux2Params}Relationships for extracting the five system calibration elements from the auxiliary vector}
\end{table}

Since the elements of $\bm{E}'_{1}$ and $\bm{J}'_{2}$ are functions of both space and frequency, our calibration procedure therefore consists of acquiring spectroscopic images of the target at a sequence of polarizer angles over the entire field of view. We then calculate the auxiliary vector and determine $\bm{E}'_{1}$ and $\bm{J}'_{2}$ for each pixel. As we will describe in the next section, the resulting maps of the five complex parameters provide complete insight into the scanner behavior and enable investigation of polarimetric samples.

\section*{Results}

\subsection*{Spatial mapping of the calibration matrix}
\begin{figure}[b!]
\centering
\includegraphics[width=6.5in]{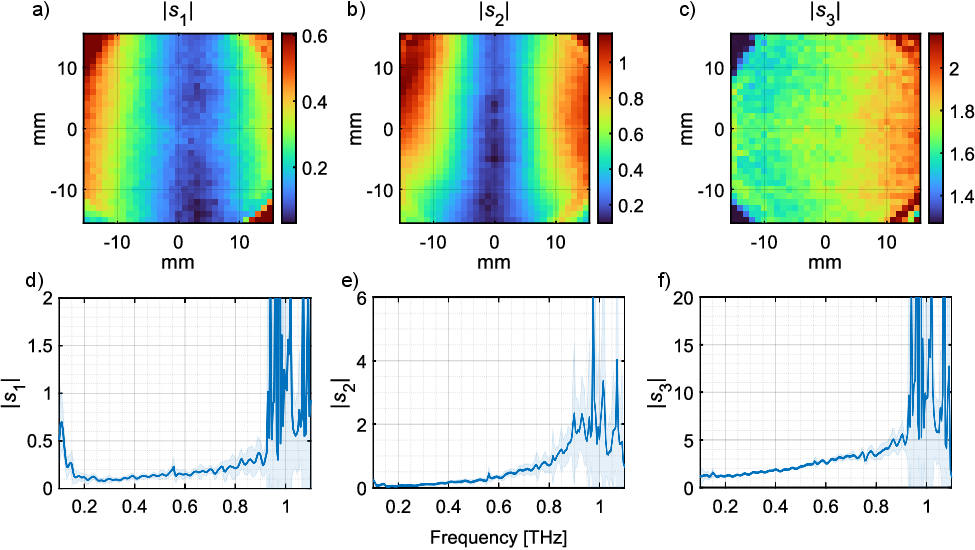}
\caption{Calibration parameters describing $\bm{J}'_{2}$, the return path through the system. (\textbf{a-c}) Spatial variation of the magnitude of $s_{1}$, $s_{2}$, and $s_{3}$, respectively, demonstrating the dependence on the beam steering position. Images show the average value over 0.25 - 0.5 THz, corresponding to the peak of the THz amplitude spectra. (\textbf{d-f}) Frequency dependence of the three parameters over a $10\times1$ mm vertical line of pixels centered at the origin. The shaded regions show one standard deviation from the mean.}
\label{fig:3J2prime}
\end{figure}

Figure~\ref{fig:3J2prime}a-c maps the spatial dependence of the elements of the calibration matrix, $\bm{J}'_{2}$, over the FOV of the scanner. Here, for simplicity, we show the magnitude of these complex quantities.
%
The parameters $|s_{1}|$ and $|s_{2}|$ in Fig.~\ref{fig:3J2prime}a and b show symmetrically increasing values as the beam is steered away from the center in the horizontal direction using the azimuthal axis of the gimballed mirror. A more subtle dependence on the elevation axis is best indicated by the the narrowing width of the blue region in the center of $s_{2}$ along the vertical axis. In comparison to the other two elements, the variation of $|s_{3}|$ in Fig.~\ref{fig:3J2prime}c is small, consisting of a slight horizontal dependence. 
%
%
Figure~\ref{fig:3J2prime}d-f show the frequency dependence of these three elements averaged over a $10\times1$ mm vertical line of pixels at the center of the image, where the polarization is not rotated by the beam steering. The normalization of $\bm{J}_{2}$ by $s_{xx}$ in Eq.~(\ref{eq:5JonesPrime}) means that any common spectroscopic or polarization-insensitive features between $E_{M,x}$ and $E_{M,y}$ are no longer captured by $\bm{J}'_{2}$, but instead included in $\bm{E}'_{1}$. For example, the attenuation due to absorption by water vapor or optical components does not appear in these plots. The near-zero values of the off-diagonal terms, $s_{1}$ and $s_{2}$, at the central area of the scan indicate that the system is successful in limiting cross-talk between the two detection channels. However, this performance metric is frequency dependent and degrades at higher frequencies, where the extinction ratio of the wire-grid polarizing beam splitter is lower\cite{microtechDatasheet}. 
In contrast to the low values of the off-diagonal terms, $|s_{3}|$ is greater than 1 over all frequencies and the full field of view, which indicates that the system is more sensitive to the $y$-polarized THz signals. 
In part, this can be attributed to the different Fresnel transmission coefficients between $s$ and $p$ polarized light passing through the silicon beam splitter. This element is a 5-mm thick wafer of high-resistivity float zone silicon mounted at a $45^\circ$ angle. The wafer is oriented such that the $s$ and $p$ components of incident light correspond to the $x$ and $y$ detection channels, respectively. Silicon has a nearly constant index of refraction over the frequency range of our system \cite{dai2004terahertz,naftaly_reference_2023}. Therefore, the beam splitter alone does not explain the frequency dependence seen in Fig.~\ref{fig:3J2prime}d-f. Furthermore, the refractive index of our wafer was measured to be $n = 3.416$, corresponding to a transmission coefficient ratio of 1.46, which does not fully account for the values of $|s_{3}|$ seen in Fig.~\ref{fig:3J2prime}c and f. Other possible contributing factors include a difference in the sensitivity of the two PCA detectors and associated electronics (i.e., transimpedance amplifiers) as well as any potential misalignment in the optical paths. Nevertheless, since this calibration is an empirical method, these effects are captured by the calibration matrix.

\begin{figure}[h]
\centering
\includegraphics[width=4.315in]{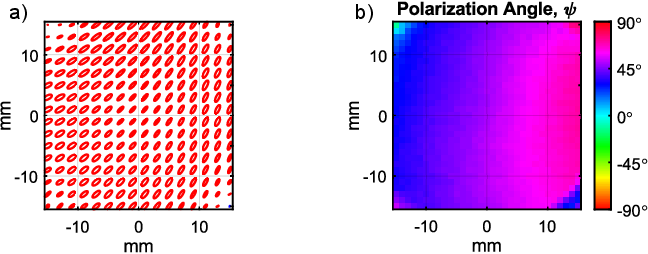}
\caption{Polarization state of THz light incident on the target. The effect of the beam steering using the gimballed mirror can be observed by the change in the angle of the polarization ellipses (\textbf{a}) and is numerically represented by the polarization angle $\psi$ in (\textbf{b}). Values represent average from 0.25 to 0.5 THz and ellipses have been averaged over $2\times2$-pixel areas for readability.}
\label{fig:4incPol}
\end{figure}

The remaining two calibration elements describe the Jones vector of the light incident on the sample, $\bm{E}'_{1}$. 
The amplitude and phase of the two orthogonal THz field components determine the shape of the polarization ellipses at the sample location, as shown in Figure~\ref{fig:4incPol}a. Over the field of view, the polarization is right-handed, with a consistent magnitude and elliptical profile, though the orientation of the ellipses rotates as the beam is scanned in the horizontal direction. This rotation is summarized by the polarization angle, $\psi$, defined as the angle between the major axis of the polarization ellipse and the $x$ axis and mapped in Figure~\ref{fig:4incPol}b. From left to right in the FOV, $\psi$ changes approximately linearly from $30^\circ$ to $70^\circ$. 
 
\begin{figure}[h]
\centering
\includegraphics[width=6.9in]{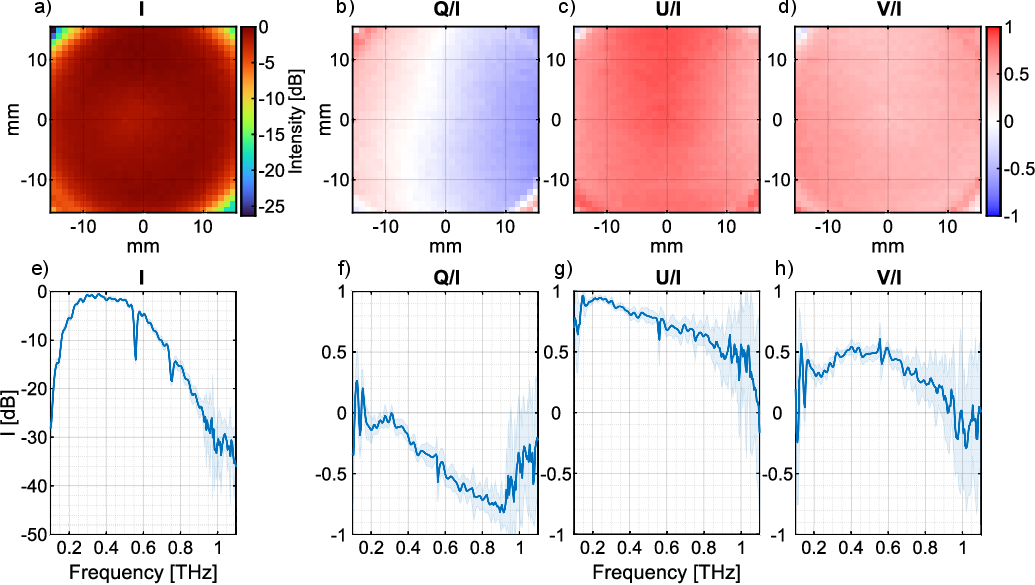}
\caption{Stokes parameters describing the field at the surface of the sample. (\textbf{a-d}) Spatial dependencies and (\textbf{e-h}) frequency dependence of the Stokes parameters of $\bm{E}'_1$. The $Q$, $U$, and $V$ values have each been normalized by the corresponding $I$ value. Images in (\textbf{a-d}) show the average value from 0.25 to 0.5 THz corresponding to the peak of the THz amplitude and the plots show the average of a $10\times1$ mm vertical line of pixels centered at the origin.}
\label{fig:5incStokes}
\end{figure}

An alternative description of the polarization state of $\bm{E}'_{1}$ can be provided by mapping the Stokes vector of the incident light beam, as shown in Fig.~\ref{fig:5incStokes}a-d. The four stokes parameters can be calculated using the relationship,
\begin{equation}
    \begin{bmatrix}
        I\\Q\\U\\V
    \end{bmatrix} = \begin{bmatrix}
        |E'_{1,x}|^{2} + |E'_{1,y}|^{2}\\
        |E'_{1,x}|^{2} - |E'_{1,y}|^{2}\\
        2\Re(E'_{1,x}E'^{*}_{1,y})\\
        -2\Im(E'_{1,x}E'^{*}_{1,y})
    \end{bmatrix}.
    \label{Eq:8Stokes}
\end{equation}
Figure~\ref{fig:5incStokes}a shows a map of the first of the four elements, $I$, which describes the total intensity of the light. The consistent magnitude across the field of view indicates how the incident power on the sample remains approximately constant despite the changing polarization state. 
The other three parameters, $Q$, $U$, and $V$, indicate the intensity in the horizontal (vertical), $+45^\circ (-45^\circ)$, and right-handed (left-handed) polarizations for positive (negative) values, respectively. Normalizing $Q$, $U$, and $V$, by $I$ then provides the degree to which the light is polarized in those components.
The spatial distribution of the normalized Stokes parameters are shown in Fig.~\ref{fig:5incStokes}b-d and provide a different interpretation to the ellipses in Fig.~\ref{fig:4incPol}. The horizontal trend of the two linear components $Q/I$ and $U/I$ show that the polarization is primarily aligned with $+45^\circ$ but transitions from a more horizontally polarized direction at the left of the FOV to a more vertically polarized direction at the right. At the same time, there is a right-handed circular polarization component with little variation across the field of view. Figure~\ref{fig:5incStokes}e shows the mean and standard deviation of $I$ for a $10\times1$ mm vertical line of pixels. 
Using our calibration procedure, the factor of $s_{xx}$ from $\bm{J}_{2}$ is included in $\bm{E}'_{1}$. This factor includes polarization-insensitive features (such as absorption lines, which were absent in Fig.~\ref{fig:3J2prime} ) as well as the spectral sensitivity of the $x$-channel (including PCA detector, transimpedance amplifier, etc.). As a result, these factors are present in $\bm{E}'_{1}$ and thus also in the Stokes vector.  
However, it can be seen from Eq.~(\ref{Eq:8Stokes}) that since $s_{xx}$ is present in both $E'_{1,x}$ and $E'_{1,y}$, it acts as a real scaling factor, $|s_{xx}|^{2}$, for the four Stokes parameters. Thus, when $Q$, $U$, and $V$ are normalized, any additional features from the detection path are eliminated. Ultimately, the frequency dependence of the normalized Stokes parameters are shown in Fig.~\ref{fig:5incStokes}f-h for a $10\times1$ mm vertical line of pixels at the center of FOV. As the frequency increases, the vertically polarized component of the light becomes more prominent ($Q/I$ approaches -1), while the $45^\circ$ and right-handed components both decrease. Thus, calibrating the scanning system is vital even for point measurements or images over a small FOV, where the spatial variation can be ignored, since this frequency dependence will still be present.

\subsection*{Validation of calibration measurements}
\begin{figure}[ht]
\centering
\includegraphics[width=6.5in]{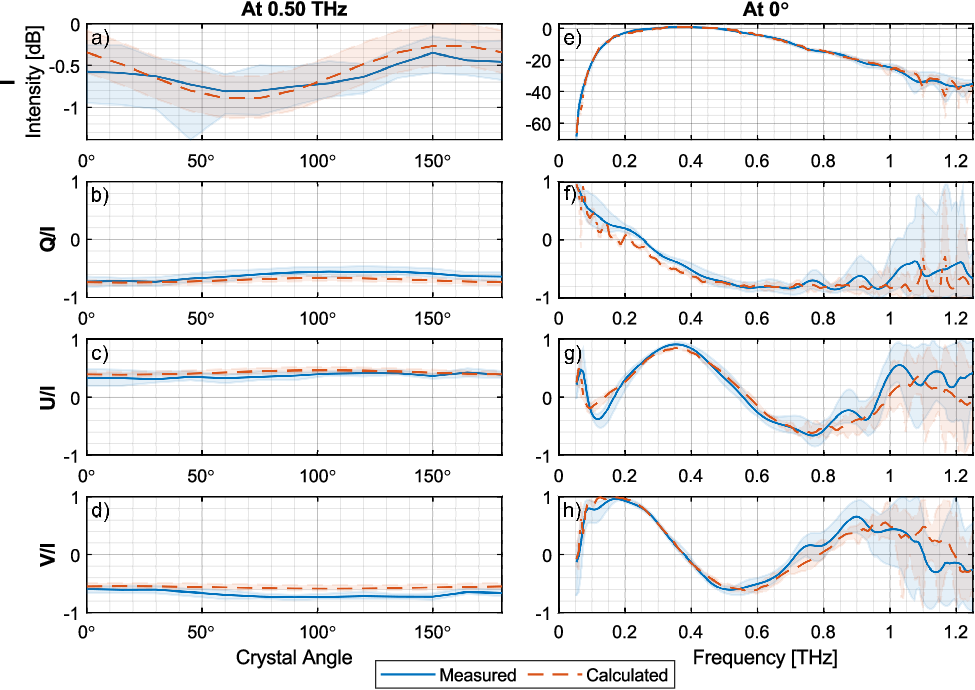}
\caption{Comparison of the measured Stokes vectors of the sapphire to the theoretically calculated values. (\textbf{a-d}) the normalized stokes parameters are shown over a range of crystal rotation angles at 0.5 THz, while (\textbf{e-h}) show the frequency dependence at $0^\circ$ rotation. The ranges shown represent the variation over a $3\times4$ mm region of pixels at the center of the crystal.}
\label{fig:6xtalStokes}
\end{figure}

In order to validate our technique, we compare measurements from a birefringent sample to those predicted theoretically using our calibration data. The crystal used here is a polished 0.52-mm-thick sapphire ($\text{Al}_2\text{O}_3$) cut orthogonal to the a-axis such that the c-axis is in the plane of the crystal face. Sapphire is chosen because it is highly birefringent and well described in the THz region \cite{grischkowsky1990far, neshat2012improved, kim2011investigation}. Due to this birefringence, the Fresnel reflection coefficients for the ordinary and extraordinary rays, $r_{o}$ and $r_{e}$, respectively, will differ even at normal incidence. For a sample crystal rotated to $\phi$ with respect to the scanning system normal direction, the sample Jones matrix representing the first surface reflection is given by\cite{mazaheri2022terahertz},
\begin{equation}
    \bm{J}_{\text{S}}(\phi) = \begin{bmatrix}
        \cos(-\phi) & \sin(-\phi)\\
        -\sin(-\phi) & \cos(-\phi)\\
    \end{bmatrix}\begin{bmatrix}
        r_{e} & 0\\
        0 & r_{o}\\
    \end{bmatrix}\begin{bmatrix}
        \cos(\phi) & \sin(\phi)\\
        -\sin(\phi) & \cos(\phi)\\
    \end{bmatrix}.
    \label{eq:9RotXtal}
\end{equation}
Using the known refractive indices of sapphire to calculate the Fresnel coefficients, we can simulate the measurement of such a crystal by substituting Eq.~(\ref{eq:9RotXtal}) into Eq.~\ref{eq:5JonesPrime}. That is, the simulated measurement $\bm{E}_{\text{calc}}(\phi)$ is given by,
\begin{equation}
    \bm{E}_{\text{calc}}(\phi) = \bm{J}'_{2}\bm{J}_{\text{S}}(\phi)\bm{E}'_{1}.
    \label{eq:10SimXtal}
\end{equation}
Figure~\ref{fig:6xtalStokes} shows the strong agreement between calculated $\bm{E}_{\text{calc}}(\phi)$ and the measured field $\bm{E}_{\text{M}}(\phi)$, expressed in terms of their normalized Stokes parameters, for a $3\times4$ mm region at the center of the crystal. 
Figure~\ref{fig:6xtalStokes}a shows how the intensity of the reflection, $I$, oscillates as the crystal is rotated at a fixed frequency (0.5 THz). A similar oscillation is also present in $Q$, $U$, and $V$ in Figure~\ref{fig:6xtalStokes}b-d such that they appear constant when normalized by $I$. On the other hand, the same measured Stokes parameters change drastically with frequency in Figure~\ref{fig:6xtalStokes}e-h. Nevertheless, there is an excellent spectral agreement between the measured Stokes parameters and the theoretical simulation across the full bandwidth of the scanner.

\section*{Discussion}

The calibration parameters described earlier are inherent to the system and robust over multiple acquisition sessions. Potential day-to-day differences in the time of arrival between the two channels can be present due to e.g., small optical miss-alignments, different ASOPS or ECOPS acquisition ranges and laser setting parameters. These shifts are compensated for by acquiring a single image of a flat mirror without the polarizer. 
Our initial alignment of the mirror and polarizer sample includes ensuring that the mirror is normal to the incident THz light. To do this, we find the orientation of the mirror, where the time of arrival (TOA) of the THz pulse across the field of view is constant for each channel. Although, it should be noted that the TOA values are not necessary the same for the two detection channels. The equivalency of normal incidence angle and constant time of arrival is provided by our telecentric $f\text{-}\theta$ scanning design\cite{harris2020design}.
For subsequent scanning sessions, measurements of the mirror at normal incidence are described by the Jones matrix equation,
\begin{equation}
    \begin{bmatrix}
        E_{\text{M},x}\\
        E_{\text{M},y}\\
    \end{bmatrix} =\begin{bmatrix}
        e^{i\theta_{x}} & 0\\
        0 & e^{i\theta_{y}}\\
    \end{bmatrix}\bm{J}'_{2}\begin{bmatrix}
        -1 & 0\\
        0 & -1\\
    \end{bmatrix}\bm{E}'_{1},
\end{equation}
where we have assumed the sample is perfectly reflecting, and thus its Jones matrix is a negative identity matrix.
The phase-shift values of $\theta_{x}$ and $\theta_{y}$ can simply be calculated by comparing the measured phase of each channel to the expected value from the previously acquired calibration set. Measurements within the new session can then be directly compared to previous measurements by including this additional phase adjustment as necessary.
Typically, the correction represents a simple time-domain shift of the form $\theta_{x/y} = 2\pi f \Delta t_{x/y}$, where $\Delta t_{x/y}$ is the difference in time of arrival between the $x$/$y$ measurements on different days\cite{jepsen_phase_2019}. As such, this measurement also accounts for the phase offset caused by session-to-session differences in reference mirror position. 
Because this phase contribution is the same for the two orthogonal components of the field, it also has no impact on the Stokes parameters. These last aspects make this adjustment valuable even for systems which do not have time-axis ambiguity between channels.


As stated in the Methods section, this calibration procedure requires the measurements of a minimum of three independent polarizer angles in order to solve for the elements of the auxiliary vector, and subsequently calculate $\bm{E}'_{1}$ and $\bm{J}'_{2}$. However, the choices of which angles to use can impact the reliability of the calibration. The periodicity of the $6\times2$ matrix in Eq.~\ref{eq:7AuxVectLabels} with respect to $\phi$ reflects the fact that the effect of the polarizer is the same under $180^\circ$ rotation, effectively limiting $\phi$ to the range $[-90^\circ, 90^\circ]$. A naive choice might include the angles $\phi = 0^\circ$, and $90^\circ$. Inspection of Eq.~(\ref{eq:6DeriveAuxVect}) shows that, for instance, at $\phi = 0^\circ$ the measured fields are given by $E_{\text{M},x} = -A = - E'_{1,x}$ and $E_{\text{M},y} = -D = -s_{2}E'_{1,x}$. At the center of the scan, where the polarization is not rotated by the system, the $E_{\text{M},y}$ component is dominated by the measurement noise, since the detection cross-talk is low. This noise is passed into the calculation of $s_{2}$. Likewise, measurements at $\phi = 90^\circ$ associate $s_{1}$ with noise. These polarizer angles therefore should be avoided when selecting which angles to use. 
The number of angles to use also influences the calibration. If just three angles are included, the resulting auxiliary vector will provide an exact solution to Eq.~\ref{eq:6DeriveAuxVect} for the measured data. However, if more than three polarizer angles are used, the system of equations becomes over-determined and solutions for the auxiliary vector will represent a best fit to the measured data. Importantly, it should be noted that even for an exact solution (i.e., when only three angles are used), the auxiliary element $E$ and the calculated value of $s_{2}E'_{1,y} + s_{3}E'_{1,x}$ are not ensured to be identical, as expected from Eq.~(\ref{eq:7AuxVectLabels}), because solving the linear system of equations for the auxiliary vector makes no explicit mathematical assumptions of the relationships between the auxiliary vector elements described by Table~\ref{tab:Aux2Params}. Additionally, imperfect extinction by the rotating polarizer or any scan-to-scan variation in $\bm{E}'_{1}$, including measurement noise, are not accounted for in Eq.~(\ref{eq:6DeriveAuxVect}). The calibration presented here is the result of measurements obtained at eight polarizer angles equally distributed between $-70^\circ$ and $70^\circ$. The auxiliary vector then was calculated to be the least-squares error solution to the measured $\bm{E}_{M}(\theta)$ using the standard matrix equation solver in MATLAB. It should be noted that regardless of how many polarizer angles are used for calibration measurements, the degrees of freedom in the fitting calculation is limited to the six parameters of the auxiliary vector. In other words, use of more measurement angles does not lead to overfitting. As a final comment, we have recently used this polarimetric version of the PHASR Scanner to map the speckles patterns induced by rough-surface samples using the Stokes parameters over the Poincar\'e sphere \cite{Xu:23}. In application such as this one, in which the objective is to investigate sample randomness or speckle patterns (in contrast to precise measurement of birefringence or other sample properties), simpler calibration procedures based on \emph{a priori} information on scanner design may be sufficient.

\section*{Conclusion}
We have presented the polarimetric version of the THz PHASR Scanner for clinical and non-destructive testing applications. We have shown the accuracy of the scanner in measuring the full spectral response of the birefringence crystal sapphire as it was rotated to various degrees. In order to accurately describes the polarization state of the light incident on the sample location and as it propagates through the scanner, we described an \emph{in situ} calibration method consisting of at least three independent measurements of a known calibration target. Our procedure is based on decomposition of the complete Jones matrix of the system and mapping the spatial and spectral dependence of its member elements over the entire FOV (1 inch) of the scanner. We showed that the telecentric beam steering approach used in this scanner can rotate the polarization angle of the incident light on the sample as a function of spatial dimensions. Finally, we showed that our approach can accurately account for the variation in the sensitivity, alignment and any potential cross-talk between the two PCA detectors used in the system. In the future, this scanner can be used to spatially map the Stokes vector of the THz light and characterize the sample response matrix in various field deployable scenarios. 
 

\bibliography{bib}

\section*{Acknowledgements}
This work was supported in part by the U.S. Army Medical Research Acquisition Activity (USAMRAA) through the Military Burn Research Program (MBRP) under Award No. W81XWH-21-1-0258, the National Institute of General Medical Sciences (GM112693), and Stony Brook University.

\section*{Author contributions statement}


K.X., M.H.A., and Z.B.H. developed the calibration method; M.H.A. and Z.B.H. conceived the experiments; K.X. and Z.B.H. conducted the experiments; Z.B.H. analyzed the results and prepared the manuscript; all authors discussed and reviewed the results and contributed to the final manuscript preparation.

\section*{Data availability statement}
The datasets used are available from the corresponding author M. H. A. on request.

\section*{Competing interests}
M.H.A and Z.B.H. disclose competing interest in patent applications 17/438,630 and PCTUS2023/022526 assigned to the Research Foundation of State University of New York. These applications cover the PHASR Scanner and high-speed data acquisition method, respectively. M.H.A. also discloses intellectual property owned by the University of Washington, US Patent No. US9295402B1\cite{arbab:16}. K.X. has no competing interests.


\end{document}